\documentclass[useAMS,usenatbib]{mn2e}
\usepackage{graphicx}
\begin{document}


%
\newcommand{\vect}[1]{\ensuremath{\mbox{\boldmath $#1$}}}
\def\APbb{ {\vect{\beta}}}
\def\APbeq{\begin{equation}}
\def\APbeqa{\begin{eqnarray}}
\def\APbeqan{\begin{eqnarray*}}
\def\APbu{ {\bf u}}
\def\APCcent{{\rm C}_{\rm cent}}
\def\APCR{{\rm C}_{\rm R}}
\def\APDR{{\rm D}_{\rm R}}
\def\APdbh{d^\bullet}
\def\APdbhn{d^\bullet_n}
\def\APDLbh{D^\bullet_L}
\def\APDLSbh{D^\bullet_{LS}}
\def\APDSbh{D^\bullet_S}
\def\APeeq{\end{equation}}
\def\APeeqa{\end{eqnarray}}
\def\APeeqan{\end{eqnarray*}}
\def\APmbh{M_\bullet}
\def\APmsun{M_\odot}
\def\APmagbht{\mu_{\rm tot}}
\def\APmagR{\mu_{\rm R}}
\def\APmagwf{\mu^{\rm wf}}
\def\APmagwft{\mu_{\rm wf}}
\def\APmagwftext{\mu_{\rm wf}^{\rm ext}}
\def\APrbh{{\rm r}_\bullet}
\def\APre{{\rm r}_{\rm E}}
\def\APrg{{\rm r}_{\rm g}}
\def\APrt{{\rm r}_{\rm tidal}}
\def\APrsun{R_\odot}
\def\APth{\theta}
\def\APthbh{\theta_\bullet}
\def\APthe{\theta_{\rm E}}
\def\APthn{\theta_n}
\def\APthcent{\Theta_{\rm cent}}
\def\APthcentext{\Theta_{\rm cent}^{\rm ext}}
\def\APthcentR{\Theta_{\rm cent}^{\rm R}}
\def\APthcentwf{\Theta_{\rm cent}^{\rm wf}}
\def\APthcentwfext{\Theta_{\rm cent}^{\rm ext, wf}}
\def\APthen{\theta^E_n}
\def\APthwf{\theta^{\rm wf}}
\def\APvth{\vartheta}
\def\APvthbh{\vartheta_\bullet}
\def\APvthen{\vartheta^E_n}
\def\APvthwf{\vartheta^{\rm wf}}


\title{On Relativistic Corrections to Microlensing Effects: Applications to the
Galactic Black Hole}
\date{2002, July 31}
\author[A. O. Petters]{A. O. Petters
\\
Duke University, Department of Mathematics, Durham, NC 27708, USA}
\pubyear{2002} \volume{000} \pagerange{1}

\maketitle \label{firstpage}
 \nokeywords

\date{Accepted XXXXXX.
      Received XXXXXXX;
 in original form XXXXXX}

\pagerange{\pageref{firstpage}--\pageref{lastpage}}
\pubyear{2002}

\maketitle

\begin{abstract}
The standard treatment of gravitational lensing
by a point mass lens $M$ is based on a 
weak-field deflection angle 
$\hat{\alpha} = 2/x_0$, where
$x_0 = r_0 c^2/2 G M$ with 
$r_0$ the
distance of closest approach to the mass of a lensed light ray.
It was shown that for a point mass lens, the 
total magnification and image centroid shift
of a point source
remain unchanged by relativistic corrections of second order
in $1/x_0$.
This paper considers these issues analytically
taking into account the relativistic
images, under three assumptions
{\bf A1}--{\bf A3}, for a Schwarzschild black hole lens with  
background point and extended sources   
having arbitrary surface brightness profiles.
The assumptions are 
{\bf A1:} The source is close to the line of sight and lies in the
asymptotically flat region outside the black hole lens;
{\bf A2:}  The observer-lens and lens-source distances are significantly
greater than the impact parameters of the lensed light rays;
and {\bf A3:} The distance of closest approach of any light ray
that does not wind around the black hole on its
travel from the source to the observer, lies in the
weak-field regime outside the black hole.
We apply our results to the Galactic black hole for
lensing scenarios where {\bf A1}--{\bf A3} hold.
We show 
that a single factor characterizes the
full relativistic correction to the weak-field image centroid
and magnification.
As  the  lens-source distance increases, the 
relativistic correction factor  
strictly decreases.  In particular,  
we find that for  point and extended sources  
about  
$10 \ {\rm pc}$ behind the 
black hole,
which is a distance significantly  
outside the tidal disruption
radius of a sun-like source, 
the relativistic correction factor
is minuscule, of order $10^{-14}$.  
Therefore,
for standard lensing configurations, any 
detectable relativistic corrections to 
microlensing by the Galactic black hole will
most likely have to come from sources significantly
closer to the
black hole.  
\end{abstract}


\section{Introduction}

Microlensing describes gravitational lensing of a source whose
multiple images are not resolved.  
Two fundamental microlensing observables are the total magnification
(photometry) and image centroid shift (astrometry) of images
of a lensed source.  These observables have important
astrophysical  applications  such as
determining the mass and distance to
the lens, angular radius of the source, etc.
(see, e.g.,
\citealt{pac96}, \citealt{pac98}, \citealt{bsvb98},
\citealt{jhp99}, \citealt{gp02}, and references therein).

A natural issue to explore is how are the 
photometry  and astrometry
of a source being lensed by
a point mass changed when the point mass is replaced
by a black hole lens.  This could have important implications
for the testability of general relativity's predictions about
how the gravitational field of a black hole affects light rays. Indeed, the
standard theoretical framework for point mass microlensing 
is based on  relativistic calculations to first-order in
$1/x_0$ about a Schwarzschild black hole,
where 
$$
x_0 = \frac{r_0}{2 \APrg}, 
\qquad \APrg = \frac{G \APmbh}{c^2},
$$ 
where $\APmbh$ is the black hole's mass and
$\APrg$ the gravitational radius.
\citet{e00} 
found that to second-order in $1/x_0$, the relativistic corrections appear
in the position and magnification of images due to
a point mass lens, while no such correction appears
in the total magnification.   \citet{lw01} also
showed that no relativistic correction to second-order in $1/x_0$
occurs for the associated image centroid shift.
In this paper, we extend the work of the previous
authors by determining an analytical expression
under assumptions {\bf A1}--{\bf A3} for the full 
Schwarzschild black hole relativistic
correction of the image centroid shift, which includes the total magnification,
for point and extended sources
with arbitrary surface brightness.

The full relativistic
corrections will then be applied
to the case of the massive black hole at the center of
our Galaxy.
Microlensing by the Galactic black hole 
has been studied by several authors
in the weak-field limit of the black hole
(e.g, \citealt{wy92}, \citealt{as99}, \citealt{al01},
\citealt{a01}).   
In addition, the precession of star
orbits 
in the strong-field regime of the black hole
was considered by,
e.g., \citet{j98a,j98b,j99}, \citet{fm00}. 
\citet{ve00} 
gave a  numerical treatment of   
the magnifications
of several relativistic images for a point source 
being lensed by a  
Schwarzschild black hole lens at the Galactic center.
They considered sources
within our Galaxy that are far away from the black hole
(about 
$8.5 \ {\rm k pc}$), while we shall
consider 
sources as close as 
$10 \ {\rm pc}$
to the black hole.
Analytical work on magnification due to
a Schwarzschild lens was also done by
several authors  for point and/or extended sources with
uniform   brightness profiles 
(e.g., \citealt{o87}, \citealt{fkn00}, \citealt{bcis01}, 
\citealt{ert02}).
As noted above, our microlensing treatment will apply not only to the magnification,
but the image centroid of extended sources
with arbitrary surface brightness.

We shall also show 
that a single factor approximates
the relativistic corrections to the
weak-field total magnification and image centroid
due to a Schwarzchild black hole lens
at the Galactic center.  The same factor applies 
to  either a
point or extended
source.
Estimates of this factor will be given 
for lens-sources distances ranging from
$10 \ {\rm pc}$ to $100 \ {\rm pc}$.
In principle, the magnification and image centroid
of sources
closer to  the Galactic black hole
should provide stronger relativistic microlensing signatures.
We shall show that even 
for sources about $10 \ {\rm pc}$ behind the black hole, 
the full relativistic
correction is still negligible.

Section~\ref{sec-image-mag} reviews 
some basic results about lensing by a Schwarzschild black hole.
In Section~\ref{sec-image-centroid}, 
we compute explicitly the image centroid due to a Schwarschild
black hole acting on point and extended sources with arbitrary brightness
profiles.  Our image centroid formula
expresses the relativistic image centroid  
in terms of the weak-field image centroid due to a point mass lens.  
Section~\ref{sec-applications} estimates the relativistic corrections
to microlensing by the Galactic black hole.

\section{Image Position and Magnification}
\label{sec-image-mag}

	We review some basic analytical results about the image positions
	and magnification due to a Schwarzschild black hole lens.

	\subsection{Lens Equation and Bending Angle}

	A Schwarzschild black hole of mass $M_\bullet$  is
	the unique static, spherically symmetric, vacuum
	(i.e., Ricci flat) spacetime that is Minkowski at
	infinity (e.g., \citealt{wld84}, p. 119):
	$$
	ds^2 = - \left(1 - \frac{\APrbh}{r}\right) c^2dt^2
	    \ + \ \left(1 - \frac{\APrbh}{r}\right)^{-1} dr^2
	  \ + \  r^2 \ d \Omega^2
	$$
	where $\APrbh$ is the 
	Schwarzschild radius
	$$
	\APrbh = 2 \APrg  = 3\ \frac{\APmbh}{\APmsun} \ {\rm km}
	$$
	and $d  \Omega^2$  the standard metric on a $2$-sphere of
	radius 1.

	Relative to
	the optical axis through the
	observer and center of the black hole, 
	let $\beta$ denote the angular position of a point source
	behind the black hole.
	Note that a
	source between the observer and
	the black hole is also lensed; in principle, 
	even the observer is lensed.
	Such situations will not be treated in this paper.
	We shall always assume the standard lensing
	configuration of a source behind the black hole

	The capture cross section $\sigma$ of
	a light ray in a Schwarzschild spacetime is
	$\sigma = \pi \ b_{\rm crit}^2,$
	where
	$b_{\rm crit}  = 3 \sqrt{3} \ \APrg$
	is the critical apparent impact parameter
	(e.g., \citealt{wld84}, p.144).
	Consequently, any light ray that makes it from the source to the observer
	will have an apparent impact parameter $b$ greater than
	$b_{\rm crit}$.   The classification of null geodesics
	in the Schwarzschild geometry divides such source-to-observer
	light rays 
	into those that loop around the black hole
	at least once before arriving at the observer and
	those that travel directly to the observer without looping
	(cf. \citealt{mtw73}, p. 674).

	Let $D_{LS}$, $D_L$, and $D_S$ be the distances
	from lens to source, observer to lens, and observer to source,
	resp.
	Denote  the angular position of a lensed image  
	relative to the line of sight by $\APvth$. 
	A negative angle $-\APvth$ 
	will correspond to an angular position
	on the side of the black hole opposite to the angular position $\APvth$.
	A trigonometric argument yields the lens equation,
	which relates 
	$\beta$ and $\APvth$ as follows 
	(\citealt{ve00}):
	\APbeq
	\label{eq-lebh}
	\tan \beta = \tan \APvth - \frac{D_{LS}}{D_S}(\tan \APvth + \tan (\hat{\alpha} - \APvth)),
	\APeeq
	where $\hat{\alpha}$ is the bending angle of the lensed ray.
	The {\it lensed images} of a light source at $\beta$ are the solutions
	$\APvth$ of the lens equation.  The {\it magnification}
	$\mu$ of a lensed image $\APvth$ is the ratio of the solid angle subtended 
	at the distance to the source by   
	the lensed image 
	to the solid angle of the unlensed source.
	Explicitly, 
	\APbeq
	\label{eq-mag}
	\mu (\APvth) = \left|\frac{\sin \beta (\APvth)}{\sin \APvth} \ 
	\frac{d \beta}{d \APvth} (\APvth) \right|^{-1}.
	\APeeq
	The lens equation cannot be solved analytically
	in closed form, so approximations appropriate to the lensing
	scenario of interest to us will 
	be employed.

	Our working assumptions are:

	\begin{itemize}

	\item[\underline{\bf A1:}] The
	source is assumed to be close to the line of sight,
	i.e.,
	$|\beta| << 1$, and lie in the asymptotically flat region outside the
	black hole, i.e., $D_{LS} >> \APrbh$.

	\item[\underline{\bf A2:}]
	In anticipation of later applications to the
	Galactic black hole, we suppose that
	$D_L >> b$ and 
	$D_{LS} >> b$, where $b$ is
	the apparent impact parameter of  
	{\it any} light ray from the source to the observer.

	\item[\underline{\bf A3:}] For any light ray 
	from the source to the observer that do not wind around the
	black hole, we shall assume that its distance of closest approach
	$r_0$ lies in the weak-field regime regime,
	i.e., $r_0 >> \APrbh$.

	\end{itemize}

	We shall now use {\bf A1} - {\bf A3} 
	to simplify
	the tangent terms in the lens equation.
	Relative to the observer,  the caustic due to the
	black hole coincides with the optical axis passing the observer
	and singularity at $r=0$.
	By {\bf A1}, the source is near to the optical axis 
	(i.e., $|\beta| << 1$),  so the source is highly magnified 
	and its angular position satisfies 
	\APbeq
	\label{eq-tan-approx-A1}
	\tan \beta \approx \beta.
	\APeeq

	In {\bf A2}, it is important to 
	add that 
	the apparent impact parameter
	$b$ and distance $r_0$ of closest approach are not equal
	in general and approximate each other
	in the weak-field limit.  In fact,
	for our lensing assumptions,
	the quantities $b$ and $r_0$ are related as follows
	(e.g., \cite{wld84}, p. 144)
	$$b = \frac{r_0}{\sqrt{1 - \APrbh/r_0}}.$$
	Consequently, we have $b > r_0$ and $b \approx r_0$ for the
	nearly flat regime $r_0 >> \APrbh$.
	Assumption
	{\bf A2} yields that $|\tan \APvth| \approx r_0/D_L << 1$, so
	\APbeq
	\label{eq-tan-approx-A2}
	\tan \APvth \approx \APvth.
	\APeeq

\begin{figure}
	\centering
	\includegraphics[height=2.2in,width=3in,]{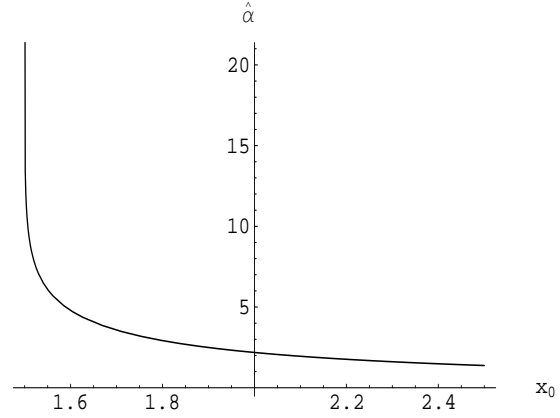}
	\medskip
	\includegraphics[height=2.2in,width=3in,]{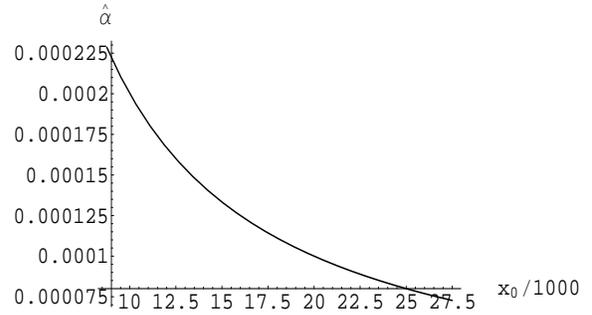}
	\caption{Bending angle $\hat{\alpha}$ (in radians) as a function
	of 
	$x_0 = r_0/\APrbh$, 
	where $r_0$ is the distance of closest approach
	to the black hole. In the top graph, we see that
	$\hat{\alpha}$
	diverges as $x_0 \rightarrow 3/2$. The bottom graph shows that 
	$\hat{\alpha}$
	is significantly less than unity for $x_0 \ga 8.75 \times 10^3$.
	In the case of the Galactic black hole,
	the range of $x_0$ in the bottom graph corresponds to 
	a distance of closest approach in the interval
	$2.29 \times 10^{-3} \  {\rm pc} \la r_0 \la 7.19 \times 10^{-3}\  {\rm pc}$.
	This yields a lens-source distance ranging from
	about $ 10 \ {\rm pc}$ to $100 \ {\rm pc}$ --- see
	Figure~\ref{fig2}.
	}
	\label{fig-1}
	\end{figure}

	Before discussing how our assumptions affect the
	term $\tan (\hat{\alpha} - \APvth)$ in (\ref{eq-lebh}),
	we consider the bending angle $\hat{\alpha}$
	under assumptions
	{\bf A1} and {\bf A2}.
	The bending angle, which is not a priori assumed to be
	small, is given by (e.g., \citealt{wnb72}, p. 189; \citealt{wld84}, p. 145):
	\APbeq
	\label{eq-bh-bending-angle}
	\hat{\alpha} (x_0) =  2 \int_{x_0}^\infty \ \frac{dx}{\sqrt{P(x)}}  - \pi,
	\APeeq
	where 
	$$P(x) = x^2 [(x/x_0)^2 (1- 1/x_0)-
		     (1- 1/x)],  \qquad x_0 = r_0/\APrbh.$$
	The integral in (\ref{eq-bh-bending-angle}) 
	is an elliptic integral and can be expressed in terms
	of an elliptic integral of the first kind.  In fact,
	set
	$$
	P(x) = \frac{1 - 1/x_0}{x^2_0} X (x)
	$$
	with 
	$$X(x) = x\left(x^3 - \frac{x_0^2}{1 - 1/x_0} (x -1)\right).$$
	The zeros of the quartic $X(x)$ 
	are
	\APbeqan
	\hspace{-0.2in}&&r_1 = x_0, \quad r_2 = \frac{x_0}{2(-1 + x_0)}
	     \left[1 - x_0 + \sqrt{-3 + 2x_0 + x^2_0}\right],\\ 
	\hspace{-0.2in}&& r_2 = \frac{x_0}{2(-1 + x_0)}
	     \left[1 - x_0 - \sqrt{-3 + 2x_0 + x^2_0}\right],
	\quad r_4 =0. 
	\APeeqan
	Note that $r_1 > r_2 >r_3 >r_4$.
	It can be shown (e.g., \citealt{hnr58}, p. 47)
	that  
	\APbeq
	\label{eq-intX}
	\int_{x_0}^\infty \frac{dx}{\sqrt{X(x)}}
	= \frac{2}{(r_1 - r_3)(r_2 - r_4)} \ F (\phi, k),
	\APeeq
	where
	$$F(\phi, k) =  \int_0^{\phi} \ (1 - k \sin^2 \phi)^{-1/2} d \phi$$
	is an elliptic integral of the first kind with arguments
	$$
	\phi = \sin^{-1}\sqrt{\frac{r_2 - r_4}{r_1 - r_4}}, 
	\qquad
	k = \frac{(r_1-r_4)(r_2 - r_3)}{(r_1 - r_3)(r_2 - r_4)}.$$
	Using (\ref{eq-intX}),
	we plot the bending angle $\hat{\alpha}$ in Figure~\ref{fig-1}.
	In addition, 
	Taylor expanding the integrand of in
	(\ref{eq-bh-bending-angle}) yields the following expansion
	for the bending angle
	(\citealt{ve00}, \citealt{ert02}):
	\APbeq
	\label{eq-expansion-bending}
	\hat{\alpha} (x_0) =  \frac{2}{x_0} \ + \
	\left(\frac{15}{16}\pi -1 \right) \ \frac{1}{x_0^2} \ + \
	{\cal O} \left(\frac{1}{x_0^3}\right).
	\APeeq

	For light rays that travel
	directly from the source to the observe without looping
	around the black hole, assumption {\bf A3} yields
	that $1/x_0 << 1$.  By (\ref{eq-expansion-bending}),
	the bending angle takes the
	standard weak-field  form:
	$$ 
	\hat{\alpha} (x_0) \approx \frac{2}{x_0}.
	$$
	Because 
	$|\hat{\alpha}| << 1$ in this case and
	$|\APvth| << 1$ by (\ref{eq-tan-approx-A2}), 
	we obtain $|\hat{\alpha} - \APvth| <<1$
	and, hence, 
	\APbeq
	\label{eq-tan-approx-A3}
	\tan(\hat{\alpha} - \APvth) \approx \hat{\alpha} - \APvth.
	\APeeq

	For the source-to-observer
	rays that do not loop, we have
	$1/x_0 \approx \APrbh/(\APvth D_L)$.
	Combined with equations
	(\ref{eq-tan-approx-A1}), (\ref{eq-tan-approx-A2}),
	and (\ref{eq-tan-approx-A3})
	we see that
	the lens equation reduces to the usual 
	{\it weak-field lens equation}:
	\APbeq
	\label{eq-wfle}
	\beta = \APvth - \frac{D_{LS}}{D_S} \ \hat{\alpha} (\APvth)
	= \APvth - \frac{\APthe^2}{\APvth},
	\APeeq
	where
	\APbeq
	\label{eq-radii-er}
	\APthe = \sqrt{\frac{2 D_{LS} \ \APrbh}{D_L \  D_S}}.
	\APeeq

	For rays that loop around
	the black hole at least once before arriving at the observer,
	the bending angle $\hat{\alpha}$
	is near a multiple of $2\pi$
	(cf. Figure~\ref{fig-1}).
	Such light rays are close to
	the unstable photon orbit at radius $3 \APrg$
	(e.g., \citealt{chn92}, p. 132).
	In this case, the bending angle has the form
	$\hat{\alpha} = 2 n \pi + \Delta \hat{\alpha}_n$,  where  
	$n\ge 1$ is the number of times the light
	ray loops around the black hole and $|\Delta \hat{\alpha}_n| << 1$
	(representing a small angular deviation from a multiple of $2 \pi$). 
	Consequently, 
	\APbeq
	\label{eq-tan-approx-sf}
	\tan (\alpha - \APvth) \approx \Delta \alpha_n - \APvth.
	\APeeq
	Equation (\ref{eq-tan-approx-sf}), along with {\bf A1} and {\bf A2}, 
	yields the following {\it strong-field lens equation} 
	for rays circling near $3 \APrg$
	(\citealt{bcis01}): 
	\APbeq
	\label{eq-sfle}
	\beta = \APvth - \frac{D_{LS}}{D_S} \Delta \alpha_n. 
	\APeeq

	We now turn to the solutions of the weak- and strong-field lens equations.

	\subsection{Einstein ring, Primary and Secondary Images}
	\label{subsec-Einsteinr-PSi}

	The solutions of the weak-field lens equation
	(\ref{eq-wfle}) are well known 
	(e.g., see \citet{PLW01}, pp. 187-191, for a detailed
	treatment).

	If $\beta = 0$ (source is on the optical axis),
	then the source is lensed
	into
	a formally infinitely
	magnified circle, called an {\it Einstein ring}, which has an angular
	radius
	$\APthe$.

	If $\beta \neq 0$ (source is off the optical axis),
	then on opposite sides of the  black hole there are 
	a primary image $\APvthwf_+$ and  secondary image
	$\APvthwf_-$ of the source:
	\APbeq
	\label{eq-imageswf}
	\APvthwf_\pm  =  \frac{\APthe}{2} \left[ (\beta/\APthe) \pm
	     \sqrt{(\beta/\APthe)^2  +  4 }\right] \  \approx \   \pm \APthe  + \frac{\beta}{2}. 
	\APeeq
	The primary and secondary images have
	magnification
	\APbeq
	\label{eq-magwf}
	\APmagwf_\pm = \frac{ (\beta/\APthe)^2 \ + \ 2}{2 |\beta/\APthe|
	    \ \sqrt{(\beta/\APthe)^2 \ + \ 4}}   \ \pm \ \frac{1}{2} \  \approx \ \frac{\APthe}{2|\beta|}.
	\APeeq 
	Note that assumption {\bf A1} was used for the approximations
	in (\ref{eq-imageswf}) and
	(\ref{eq-magwf}).
	The total magnification is
	$$\APmagwft = \APmagwf_+ + \APmagwf_- = \frac{\APthe}{|\beta|}.$$

	To simplify expressions like (\ref{eq-imageswf})and 
	(\ref{eq-magwf}) that involve $\APthe$,
	we shall scale all angles 
	as follows:
	$$u = \frac{\beta}{\APthe}, \qquad \APth = \frac{\APvth}{\APthe},$$
	The weak-field image positions and magnifications now become
	\APbeq
	\label{eq-images-mag-wf-s}
	\APthwf_\pm =  \frac{u}{2} \ \pm 1,
	\qquad
	\APmagwf_\pm \  = \  \frac{1}{2|u|},
	\qquad \APmagwft =  \frac{1}{|u|}.
	\APeeq
	Note that the angular image positions
	and magnifications
	in (\ref{eq-images-mag-wf-s}) 
	for the weak field regime
	coincide with that obtained from
	lensing by a point mass.

	\subsection{Relativistic Einstein rings}

	If $\beta =0$, then
	the source is lensed into the formally infinitely
	magnified Einstein ring mentioned in Section~\ref{subsec-Einsteinr-PSi},
	along with 
	an infinite sequence of relativistic Einstein
	rings near $3 \APrg$.  
	The angular radii $\APvthen$ of the relativistic Einstein rings are
	given by the solutions of 
	the strong-field lens equation (\ref{eq-sfle}) for $\beta = 0$
	(\citealt{bcis01}):
	\APbeq
	\label{eq-radiin-er}
	\APvthen = \APvthbh \left(A_0 \ e^{-(2n +1)\pi}
	       + B_0\right),
	\APeeq
	where
	$$
	\APvthbh  =  \frac{\APrbh}{D_L}, \quad
	A_0   = \sqrt{3} \left(\frac{18}{2 + \sqrt{3}} \right)^2,
	\quad B_0   = \frac{3 \sqrt{3}}{2}.
	$$
	Note that $\APvthbh$
	is approximately the angle spanned by the Schwarzschild radius.
	For simplicity, set
	$$\APdbhn \  = \  A_0 \ \APvthbh \ 
	\ e^{-(2n +1)\pi}$$
	and rewrite $\APvthen$ as
	$$\APvthen = \APdbhn \  + \  \APvthbh \ B_0.$$
	Since 
	$\APvthbh << 1,$
	we have
	$\APdbh_1 <  1$  (because
	$A_0 \ e^{-3 \pi} \approx 3.25 \times 10^{-3}$).  Consequently, 
	each term in the strictly decreasing sequence $\{\APdbhn\}_{n =1}^\infty$ 
	obeys
	\APbeq
	\label{eq-dn-1}
	\hspace{1in}   \APdbhn \le \APdbh_1 <  1, \qquad n=1,2,\dots.
	\APeeq

	\subsection{Relativistic Images}

	If $\beta \neq 0$,
	then we know from Section~\ref{subsec-Einsteinr-PSi}
	that on opposite sides of the  black hole there are 
	a primary image $\APvthwf_+$ and  secondary image
	$\APvthwf_-$ from light rays that travel directly
	(no looping around the black hole)
	to the observer.
	In addition, there  are 
	two infinite sequences (on opposite sides of the black hole)
	of  relativistic images
	$\APvth^\pm_n$
	that are due to light rays looping $n$ times
	around the black hole near $3 \APrg$.
	The following expressions for the solutions  $\APvth^\pm_n$ of  
	the strong-field lens equation (\ref{eq-sfle})
	are convenient (\citealt{bcis01}):
	\APbeq
	\label{eq-thRn}
	\APvth^\pm_n  =  \pm \APvth^E_n \ + \ \frac{D_S}{D_{LS}} \ \APdbhn \ \beta.
	\qquad n =1,2,\dots.
	\APeeq
	The images $\APvth^+_n$  (resp.,  $\APvth^-_n$) result from $n$ clockwise 
	(resp., counter-clockwise) windings around
	the black hole.

	Using (\ref{eq-mag}), the magnifications of the relativistic images can be shown
	to be approximately as follows (\citealt{bcis01}, \citealt{ert02}):
	\APbeq
	\label{eq-magRn}
	\mu^\pm_n  
	= \ \frac{D_S}{D_{LS}} \APvthen \ \APdbhn \ \frac{1}{|\beta|}.
	\APeeq
	The total magnification is then
	$$
	\APmagR  =  
	\sum_{n=1}^\infty \mu^+_n \ + \ \sum_{n=1}^\infty \mu^-_n.
	$$
	Equations (\ref{eq-radiin-er}) and (\ref{eq-magRn}) imply that 
	\APbeq
	\label{eq-mag-R-1}
	\APmagR =  
	\frac{2 \ D_S}{|\beta| \ D_{LS}} 
	 \left[ \sum_{n=1}^\infty (\APdbhn)^2 \ + \ 
	B_0 \ \APvthbh \  \sum_{n=1}^\infty \APdbhn
	\right].
	\APeeq
	By (\ref{eq-dn-1}),
	both series in  (\ref{eq-mag-R-1}) are geometric series.
	Hence,
	\APbeq
	\label{eq-mag-R-2}
	\APmagR =
	\frac{2 \ D_S \ \APvthbh^2}{D_{LS}} \ \frac{C_0}{|\beta|},
	\APeeq
	where
	\APbeq
	\label{eq-C0}
	C_0   =  
	    A_0 
	\left[ A_0  \ 
	 \frac{e^{-6 \pi}}{1- e^{-4 \pi}} 
	\ + \ B_0 \  \frac{e^{-3 \pi}}{1- e^{-2 \pi}}
	\right].
	\APeeq
	Equations (\ref{eq-thRn}),
	(\ref{eq-magRn}), 
	and  (\ref{eq-mag-R-2}) simplify to
	\APbeqa
	\label{eq-thRn-s}
	\APth^\pm_n & =&  \pm \APth^E_n \ + \ \frac{D_S}{D_{LS}} \ \APdbhn \ u,
	\qquad  \APth^E_n = \frac{\APvth^E_n}{\APthe},
	\\
	\label{eq-magRn-s}
	\mu^\pm_n  & = & \frac{D_S}{D_{LS}} \APthen \ \APdbhn \  \frac{1}{|u|},
	\\
	\label{eq-mag-R-s}
	\APmagR &=&
	\frac{2\ D_S \ \APvthbh^2}{D_{LS}\ \APthe} \ \frac{C_0}{|u|}.
	\APeeqa
	By (\ref{eq-mag-R-s}), the relativistic and total
	magnifications for a point source are then given
	as follows
	(\citealt{bcis01}, \citealt{ert02}):
	\APbeq
	\label{eq-magbht}
	\APmagR = \APCR \ \APmagwft, \qquad
	\APmagbht = \APmagwft + \mu_R =
	\left(1 + \APCR \right) \ \APmagwft,
	\APeeq
	where
	\APbeq
	\label{eq-CR}
	\APCR =
	\frac{ 2 D_S \ \APvthbh^2 \ C_0}{D_{LS}\ \APthe}
	= 4 \ C_0 \ \APthbh^3, \qquad \qquad
	\APthbh = \frac{\APvthbh}{\APthe}.
	\APeeq
	Equation (\ref{eq-magbht}) shows that
	$\APCR$ determines the relativistic correction
	to the total weak-field magnification.

	We also refer interested readers to the elegant generalization
	by \citet{ert02} of equations 
	(\ref{eq-thRn-s})---(\ref{eq-mag-R-s})
	to the case of a  Reissner-Nordstr\"om black hole lens.

	\section{Image Centroid}
	\label{sec-image-centroid}

In this section, we derive a formula using {\bf A1}--{\bf A3}
that expresses the image centroid 
of a Schwarzschild black hole as a relativistic correction 
to the weak-field image centroid of 
a point mass lens.

	\subsection{Image Centroid Formula}

	The center-of-light or centroid of all the images
	of a source at $u$ is 
	\APbeq
	\label{eq-cent-def}
	\APthcent  = \frac{
	\displaystyle 
	\left(\mu_+ \ \APth_+  \ + \ \mu_- \ \APth_-\right)
	   + \sum_{n=1}^\infty \left(\mu_n^+ \ \APthn^+  \ + \ \mu_n^- \ \APthn^-\right)
	}{\APmagbht},
	\APeeq
	where $\APmagbht$ is the total magnification.
	Since the centroid shift is 
	$$\delta \APthcent = \APthcent -u,$$
	it suffices to study $\APthcent$.

	We now express the image centroid as
	\APbeq
	\label{eq-cent}
	\APthcent  = \frac{\APmagwft}{\APmagbht} \ \APthcentwf
	 \ + \ \frac{\APmagR}{\APmagbht} \ \APthcentR,
	\APeeq
	where $\APthcentwf$ and $\APthcentR$ are respectively the centroids
	of the weak-field and relativistic images.
	Explicitly,
	$$
	\APthcentwf = \frac{\APmagwf_+ \ \APthwf_+  \ + \ \APmagwf_- \ \APthwf_-}{\APmagwft}
	$$
	and
	$$
	\APthcentR = \frac{\sum_{n=1}^\infty \left(\mu_n^+ \ \APthn^+  
	\ + \ \mu_n^- \ \APthn^-\right)}{\mu_R}.
	$$ 
	By (\ref{eq-images-mag-wf-s}) and (\ref{eq-mag-R-s}),
	the weak-field image centroid is 
	\APbeq
	\label{eq-cent-wf}
	\APthcentwf = \frac{u}{2}.
	\APeeq
	Recall that (\ref{eq-cent-wf}) is the weak-field image centroid for
	source positions near the caustic.  Indeed, the full shifted image 
	centroid due to a point mass lens is an ellipse
	(e.g., \citealt{hnp95},\citealt{my95},\citealt{w95}).

	Using (\ref{eq-thRn-s}) and (\ref{eq-magRn-s}),
	we get
	$$
	\APthcentR =  \frac{2}{\APmagR} \ \left(\frac{D_S}{D_{LS}}\right)^2
	  \ \frac{u}{|u|}  \sum_{n=1}^\infty  \ \APthn^E \ (\APdbhn)^2.
	$$
	Now, observe that
	\APbeq
	\label{eq-thndn}
	\sum_{n=1}^\infty \ \APthn^E \ (\APdbhn)^k
	= \frac{C_0 (k)\ \APvthbh^{k+1}}{\APthe },
	\qquad k =1,2,\dots,
	\APeeq
	where
	\APbeq
	\label{eq-C0k}
	C_0 (k)   =
	   A^k_0
	\left[ A_0  \ E (k+1) \ + \ B_0 \ E(k)\right]
	\APeeq
	with
	$$E(k) = \frac{e^{-3k\pi}}{1- e^{-2k\pi}}.$$
	Note that $C_0(1) = C_0$.
	Equation (\ref{eq-thndn}) yields
	$$
	\APthcentR =  \frac{2}{\APmagR} \ \left(\frac{D_S}{D_{LS}}\right)^2
	\ \frac{C_0 (2) \ \APvthbh^3}{\APthe}  \ \frac{u}{|u|}  
	=
	\frac{8 \ C_0 (2)\ \APthbh^5}{\APmagR} \ \frac{u}{|u|}.
$$
Equations (\ref{eq-magbht}), (\ref{eq-CR}),
and (\ref{eq-cent-wf})
imply
\APbeq
\label{eq-cent-R}
\APthcentR =  \APDR \ \APthcentwf,
\APeeq
where
\APbeq
\label{eq-DR}
\APDR = 4 \ \frac{C_0 (2)}{C_0 (1)} \ \APthbh^2.
\APeeq
Hence, the image centroid (\ref{eq-cent}) simplifies to 
\APbeq
\label{eq-cent-main}
\APthcent  = \APCcent \ 
         \APthcentwf,
\APeeq
where
\APbeq
\label{eq-Ccent}
\APCcent = \frac{1 \ + \ \APDR \ \APCR}{1 \ + \ \APCR}.
\APeeq
It follows that 
the constants  $\APCR$ 
and $\APDR$ determine the full relativistic 
correction to the image centroid 
due to a point mass lensing
a source near the caustic.

\subsection{Extended Source}

The total magnification of an extended source $D$ with
surface brightness $S$ is given by
$$
\mu^{\rm ext}_{\rm tot} 
= \frac{\int_D d \APbu \ S (\APbu) \APmagbht (\APbu)}
    {\int_D d \APbu \ S (\APbu)},
$$
where $\APbu = u \ \hat{\APbb}$ with 
$\hat{\APbb} = \APbb/|\APbb|$,  the unit angular vector position of the source.
The spherical symmetry of the Schwarzschild black hole implies that
$\APmagbht (\APbu) = \APmagbht (u)$.  
Consequently, equation (\ref{eq-magbht}) yields
\APbeq
\label{eq-magtot-ext}
\APmagbht^{\rm ext} = \left(1 \ + \ \APCR \right) \ \APmagwftext,
\APeeq
where
$$
\APmagwftext
=  \frac{\int_D d \APbu \ S (\APbu) \APmagwft (u)}
    {\int_D d \APbu \ S (\APbu)}
$$
is the weak-field extended source total magnification.
Readers are referred to \citet{wm94} for a discussion
of $\APmagwftext$ for a point mass lens.

The image centroid of the extended source is
$$
\APthcentext  = \frac{\int_D d \APbu \ S (\APbu) \ \APthcent \ \APmagbht (\APbu)}
    {\int_D d \APbu \ S (\APbu)\  \APmagbht (\APbu)}.
$$
Equation (\ref{eq-cent-main}) gives 
\APbeq
\label{eq-cent-main-ext}
\APthcentext  = \APCcent \
         \APthcentwfext,
\APeeq
where
$$
\APthcentwfext
=  \frac{\int_D d \APbu \ S (\APbu) \ \APthcentwf (u) \ \APmagwft (u)}
    {\int_D d \APbu \ S (\APbu)\  \APmagwft (u)}
$$
is the weak-field extended image centroid
--- see \cite{mw98} for a treatment of $\APthcentwfext$
with a point mass lens.
Thus, as in the point source case, the quantities
$\APCR$ and $\APDR$ determine the relativistic corrections
to the extended source image centroid, as well as the
total magnification, for a point mass lens.

\section{Applications to the Galactic Black Hole Lens}
\label{sec-applications}

In this section, we estimate the relativistic
corrections to the weak-field total magnification and
image centroid for the case of the Galactic black hole 
under {\bf A1}--{\bf A3}.

We begin with estimates of some needed quantities.
The constants $C_0 (1)$ and $C_0 (2)$ are given approximately
as follows:
$$
C_0 (1) \approx 8.47 \times 10^{-3},
\qquad
C_0 (2) \approx 2.75 \times 10^{-5}.
$$
For the massive black hole at the Galactic center, we have
(\citealt{g98}):
$$\APmbh \approx 2.6\times 10^6 \APmsun, \qquad D_L \approx 8.5 \ {\rm kpc}.$$
The linear and angular Schwarzschild radii of the black hole are
\APbeq
\label{eq-gbh-r-vth}
\APrbh \approx 5.23 \times 10^{-2} \ {\rm AU},
\qquad \APvthbh = \frac{\APrbh}{D_L}
\approx  6.46 \ {\rm \mu as}.
\APeeq
Since the relativistic images are near 
$3 \APrbh/2$, those on opposite sides of the
black hole have angular spacings
of order  $ 3 \APvthbh \approx 19.38 \ {\rm \mu as}$,
which is outside the resolving capabilities of near-future instruments.
The expected resolution of the Space Interferometry Mission 
to be launched in 2009 is about
$10 \ {\rm mas}$.  Hence, the source is microlensed by
the strong-field region near the black hole.

\begin{figure}
\centering
\includegraphics[height=2.2in,width=3in]{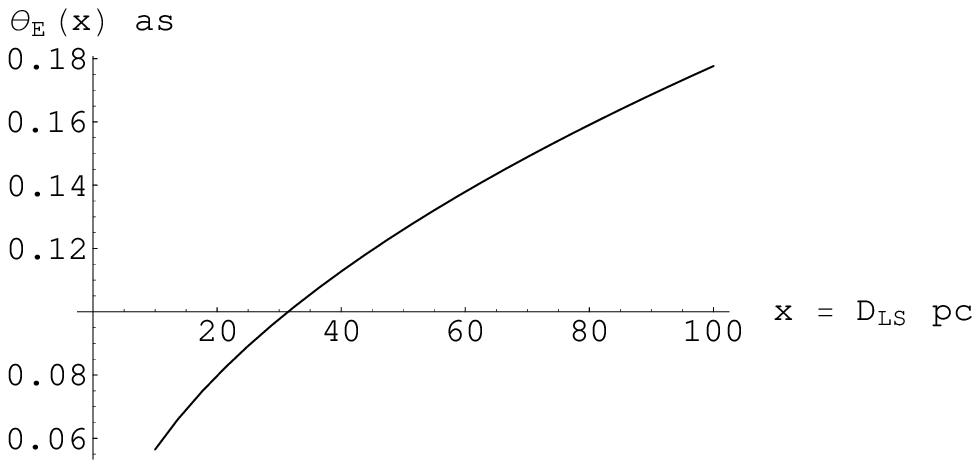}
\medskip
\includegraphics[height=2.2in,width=3in]{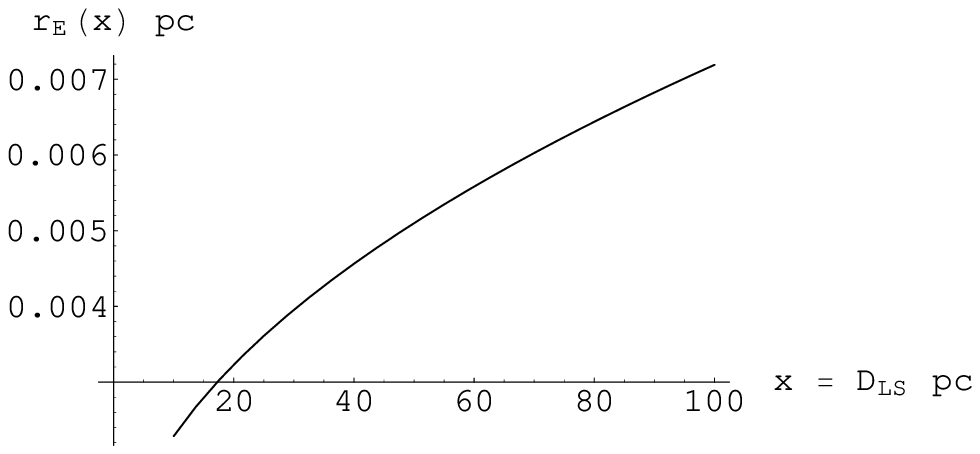}
\caption{The angular (top graph) and linear (bottom graph)
Einstein ring radii as a function
of lens-source distance in the range  $10 \ {\rm pc} \le D_{LS} \le 100 \ {\rm pc}.$
For these distances, the linear Einstein ring radius $\APre$ approximates the
distance $r_0$ of closest approach as well as the impact parameter $b$
(since $r_0 \approx b$).  The value of $\APre$ is approximately
$2.29 \times 10^{-3}  \ {\rm pc}$ for $D_{LS} \approx 10 \ {\rm pc}$.
}
\label{fig2}
\end{figure}

The (weak-field) angular Einstein
radius as a function of the lens-source distance
$x = D_{LS}$
is given by 
\APbeq
\label{eq-the}
\APthe (x)
= \left(\frac{2 \APvthbh\ x}{D_L + x}\right)^{1/2}
\approx 1.65  \sqrt{ \frac{x}{D_L +  x}}\  \ {\rm as}.
\APeeq
The function $\APthe (x)$ has a positive derivative and so is
strictly increasing  --- see
Figure~\ref{fig2}.  
The angular diameter of 
the Einstein ring is approximately the angular
spacing between the primary and secondary images.
For 
$10 \ {\rm pc} \le x \le 100 \ {\rm pc},$
the range of the angular spacing
is 
$
0.1 \ {\rm as}  \la 2\APthe (x) \la 0.4 \ {\rm as},
$
which is at the boundary of the resolving capabilities of
present day instruments.

We also have 
$$
\APthbh (x) = \frac{\APvthbh}{\APthe (x)}
\approx 3.922 \times 10^{-6} \ \sqrt{\frac{D_L +  x}{x}} 
$$
and by (\ref{eq-DR}) we get
\APbeq
\label{eq-DR-est}
\APDR (x) \approx  2 \times 10^{-13} \left(\frac{D_L +  x}{x}\right).
\APeeq

\begin{figure}
\centering
\includegraphics[height=2.2in,width=3in]{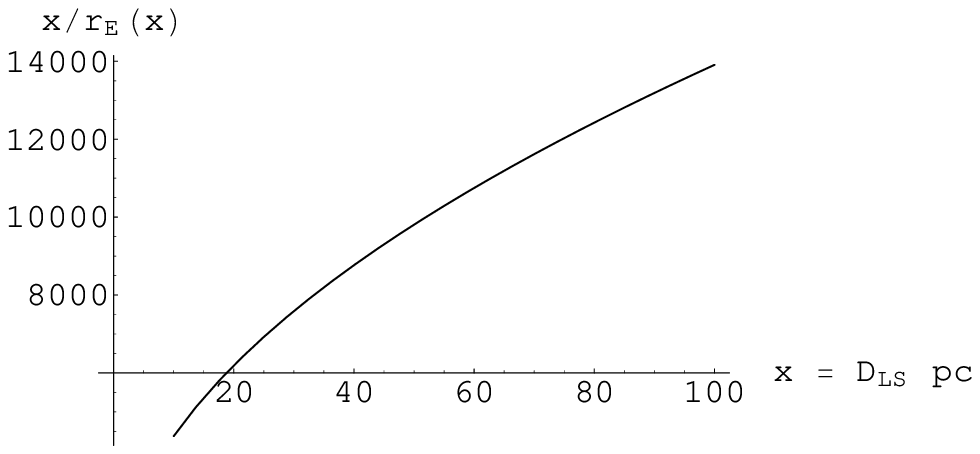}
\medskip
\includegraphics[height=2.2in,width=3in]{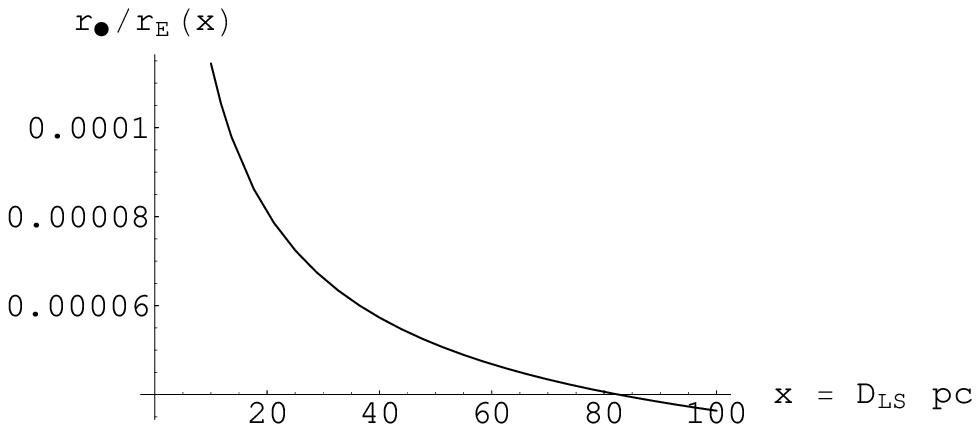}
\caption{The top graph shows that $D_{LS} >> \APre$ for
$D_{LS} \ge 10 \ {\rm pc}$, while the bottom illustrates
that $\APrbh/\APre << 1$.  The latter implies that
the primary and secondary images lie in the nearly flat
region outside the Galactic black hole. 
}
\label{fig3}
\end{figure}

Let us now verify that assumptions {\bf A1}--{\bf A3} hold.
We are supposing that $|\beta| << 1$ and $D_{LS} \ge 10 \ {\rm pc}$,
which by (\ref{eq-gbh-r-vth}) yields 
$D_{LS}/\APrbh  \approx  3.82 \times 10^7 >> 1,$ 
so assumption {\bf A1} is satisfied.
Now, the largest value $b_{\rm max}$ of the impact parameter of light rays from the
source to the observed occurs for the rays that travel to the observer
without looping around the black hole.  These rays passing through
the nearly flat region outside the black hole, which means that
the associated impact parameter and distance of closest approach
are approximately
equal.  Equation (\ref{eq-imageswf}) shows that the angular positions of the 
closest approach of the light rays are given 
approximately by $\pm \APthe$ (since $|\beta| << 1$).
It follows that $b_{\rm max}$ 
can be approximated
by the linear Einstein ring radius,
$\APre (x) = D_L \APthe (x)$ (Figure~\ref{fig2}).
The function $x/\APre (x)$ is strictly increasing
with a minimum value of approximately 
$ x/\APre (x) \ga 4.38 \times 10^3 >>1$
for $x \ge 10 {\rm pc}$
--- see Figure~\ref{fig3}.
In addition, 
since $\APre (x)$ is also  strictly increasing 
with maximum 
$\APre^{\rm max} \approx \ 8 \times 10^{-6} \ D_L$ for
$x >> D_L$
(see (\ref{eq-the})), it follows that
$D_L/\APre \ge D_L/\APre^{\rm max} \approx 1.3 \times 10^5 >> 1.$
Consequently, assumption {\bf A2} holds.
Now, 
since $\APrbh/\APre (x)$ is strictly decreasing
with a maximum value of order $10^{-4}$ for
$x \ga 10  \  {\rm pc}$, the primary and secondary
images lie in the nearly flat region outside the
black hole.  In particular, assumption {\bf A3} is also satisfied.

We add that since 
$\APrbh/x$ is at most of order $10^{-8}$ for
$x \ge 10  \ {\rm pc}$, a source at distance $x$ lies in a region
of space that is ``flatter''
than the region  
near the linear Einstein radius $\APre (x)$
(since $\APrbh/\APre (x) \la 10^{-4}$).  
Furthermore,
for a source star of
mass $M_*$ and radius $R_*$, the closest it can get
to the Galactic black hole without being
thorn apart is 
given roughly by the tidal disruption radius of the source
(e.g., \citealt{mt99}):
$$
\APrt  
  \approx 137.5 \ \left(\frac{\eta^2_* \ \APmbh}{2.6 \times 10^6 \ \APmsun}\right)^{1/3}
              \  \left(\frac{\APmsun}{M_*}\right)^{1/3}
                \     R_*, 
$$
where $\eta_*$ is a parameter of order unity that
depends on the model of the source.
For $M_* \approx  \APmsun$ and
$r_* \approx \APrsun$, the tidal radius
is then of order
$$
\APrt  \approx 137.5 \ \APrsun \approx 0.64 \ {\rm AU}. 
$$
Hence,  a source at  $x \ga 10 \ {\rm pc}$  is significantly  outside 
its tidal disruption radius.

\begin{figure}
\centering
\includegraphics[height=2.2in,width=3in]{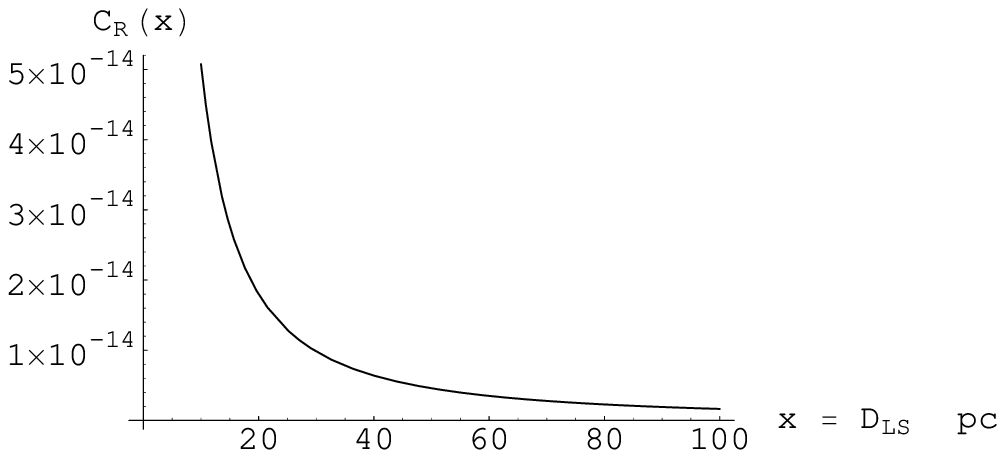}
\medskip
\includegraphics[height=2.2in,width=3in]{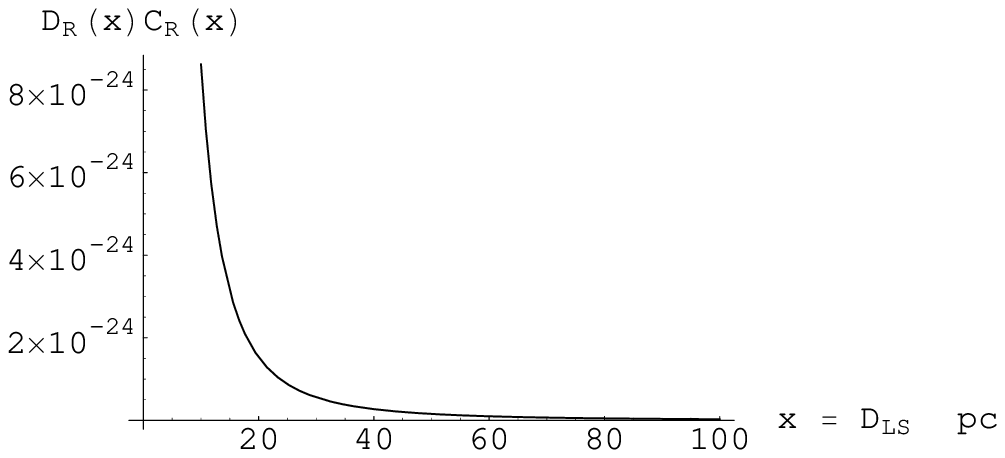}
\caption{The quantities $\APCR$ and $\APDR \APCR$ 
as a function
of the distance $D_{LS}$
of the source from the
Galactic black hole, where
$10 \ {\rm pc} \le D_{LS} \le 100 \ {\rm pc}$.
Note that these distances are 
significantly outside the tidal disruption radius,
$\APrt \approx 0.64 \ {\rm AU}$,
for a sun-like source.
The bottom figure guarantees that $\APCR$ describes
the relativistic correction to both the weak-field image
centroid and total magnification (see discussion in text).
The correction
is clearly  
negligible for
sources $D_{LS} \ga 10 \ {\rm pc}$.
}
\label{fig4}
\end{figure}

For the Galactic black hole,
equation (\ref{eq-CR}) shows that  the
relativistic correction to the weak-field total magnification
is 
\APbeq
\label{eq-CR-est}
\APCR (x)
\approx  2.1 \times 10^{-18} \ \left(\frac{ D_L +  x}{x}\right)^{3/2}.
\APeeq
Since the derivative of $\APCR (x)$ 
is  always negative,
the correction $\APCR (x)$ is a strictly decreasing
function of the lens-source distance $x$   --- see
Figure~\ref{fig4}. 
The figure  shows that  the relativistic correction $\APCR$
to the magnification is at most of order $5 \times 10^{-14}$ for
sources with $D_{LS} \ga 10 \ {\rm pc}$.

Turning to the relativistic image centroid, 
equations (\ref{eq-DR-est}) and (\ref{eq-CR-est})
yield 
that 
the derivative of $\APDR (x)  \APCR (x)$ is
always negative.  Consequently, the
function $\APDR (x) \APCR (x)$ is strictly decreasing with increasing
lens-source distance --- Figure~\ref{fig4}.  
Furthermore, the figure yields that 
$\APCR (x)$ is at most of order $10^{-14}$ and
$\APDR (x)  \ \APCR (x)$ at most of order $10^{-24}$ for
$x \ga 10 \ {\rm pc}$, so
the relativistic image centroid
factor (\ref{eq-Ccent}) can be approximated as follows:
\APbeq
\label{eq-Ccent-est}
\APCcent (x) \approx 1 - \APCR (x)  \qquad {\rm for} \ x \ga 10 \ {\rm pc}.
\APeeq
It follows that 
the relativistic correction $\APCR$ to the weak-field total magnification
also serves as a relativistic correction to the weak-field
image centroid.
The value $\APCR (x) =1$ corresponds to the weak-field case,
which clearly holds approximately 
for 
$x \ga 10 \ {\rm pc}$.
In light of Figure~\ref{fig4}, 
measurable 
relativistic corrections to the
total magnification and image centroid due to standard
gravitational lensing by the Galactic
black hole would most likely require 
sources very close to the
black hole.

\section{Conclusion}

Previous work on gravitational lensing by the
black hole at the Galactic center investigated the relativistic
corrections to the weak-field total magnification and
image
centroid to second order in 
$1/x_0 = 2 G M/(r_0 c^2)$, where  
$r_0$ is the
distance of closest approach of the light ray to the black hole.
It was shown recently that for a point mass lens the 
total magnification and image centroid shift
of a point source
remain unchanged by relativistic corrections of second order
in $1/x_0$.
We computed the relativistic corrections 
for a Schwarzschild black hole lens under assumptions
{\bf A1}--{\bf A3}.   These corrections
were applied to the case of the massive black hole at the
Galactic center.
We found that the weak-field magnification and
image centroid have approximately the same relativistic correction.
This correction is a strictly decreasing
function of the lens-source distance $D_{LS}$. 
For 
$D_{LS} \ge 10 \ {\rm pc}$, the relativistic correction
is of order at most $10^{-14},$ 
a minuscule correction.
Hence, for standard lensing configurations,
a nontrivial relativistic
correction to microlensing by the Galactic black hole
would likely have to come from sources 
deep inside
the black hole's potential well.

\section*{Acknowledgments}

Many thanks to Scott Gaudi for stimulating  discussions and
the referee for a helpful clarifying comment.
This work was supported in part by an Alfred P. Sloan Research Fellowship
and NSF Career grant DMS-98-96274.

\label{lastpage}

\end{document}